\documentclass[prb,twocolumn,amsmath,amssymb,superscriptaddress,showpacs]{revtex4-1}
\usepackage{graphicx}
\usepackage{dcolumn}
\usepackage{multirow}
\usepackage{bm}
\usepackage{ulem}
\usepackage{color}

\begin{document}

\author{J\"orn W. F. Venderbos} 
\affiliation{Department of Physics, Massachusetts Institute of Technology, Cambridge, Massachusetts 02139, USA}

\author{Marco Manzardo}
\affiliation{Institute for Theoretical Solid State Physics, IFW Dresden,
Helmholtzstr. 20, 01069 Dresden, Germany}

\author{Dmitry V.  Efremov} 
\affiliation{Institute for Theoretical Solid State Physics, IFW Dresden,
Helmholtzstr. 20, 01069 Dresden, Germany}

\author{Jeroen van den Brink}
\affiliation{Institute for Theoretical Solid State Physics, IFW Dresden,
Helmholtzstr. 20, 01069 Dresden, Germany}
\affiliation{Department of Physics, Technical University Dresden, D-1062 Dresden, Germany}

\author{Carmine  Ortix}
\affiliation{Institute for Theoretical Solid State Physics, IFW Dresden,
Helmholtzstr. 20, 01069 Dresden, Germany}
\affiliation{Institute for Theoretical Physics, Center for Extreme Matter and Emergent Phenomena, Utrecht University, Leuvenlaan 4, 3584 CE Utrecht, The Netherlands}

\title{Engineering interaction-induced topological insulators in a $\sqrt{3} \times \sqrt{3}$ substrate-induced honeycomb superlattice}

\begin{abstract}
\noindent
We consider a system of spinless fermions on the honeycomb lattice
with 
substrate-induced 
modulated electrostatic potentials  tripling the unit cell. 
The resulting non-Abelian ${\cal SU} (2)$ gauge fields act cooperatively to realize a quadratic band crossing point (QBCP). 
Using a combination of mean-field theory and renormalization group techniques, we show that in the QBCP regime, arbitrarily weak repulsive electronic interactions drive the system into the quantum anomalous Hall state.
This proves
that substrate-induced local voltages are an effective knob to induce the spontaneous formation of a topological quantum phase.
\end{abstract}

\date{\today}
\maketitle

\section{Introduction}
Realizing topologically non-trivial states of matter in band insulators has been the subject of growing interest in recent years. In the quantum anomalous Hall (QAH) insulator \cite{hal88}, the time-reversal symmetry broken ground state has a bulk insulating gap but has topologically protected chiral edge states. In the time-reversal invariant quantum spin-Hall (QSH) insulator  \cite{kan05b,kan05,ber06,kon07} a pair of helical edge states, with electrons of opposite spin counterpropagating at the sample boundaries, are mandated by the non-trivial topology of the bulk electronic states. 
The behavior of non-interacting insulating topological phases is presently well understood \cite{fu07,fu07b,moo07,has10}. 
Taking into account the effect of electronic correlations, many intriguing questions arise. 
For instance, electronic interactions may give rise to insulating topological phases without non-interacting analogs, i.e. the symmetry protected topological phase~\cite{sent14}, or fractional topological insulators~\cite{ftirefs1,ftirefs2}. 
Another class of interacting topological states are phases of interacting electrons in which chiral orbital currents or spin-orbit coupling are spontaneously generated by electron correlations. In these quantum states conventional symmetry breaking order is inextricably linked to their nontrivial topological character, and they have been called topological Mott insulators (TMI)~\cite{rag08}. 

Both the QAH and the QSH insulator were originally conceived in the context of honeycomb lattice Dirac fermions~\cite{hal88,kan05b}, by adding spin (in)dependent terms to the Dirac Hamiltonian 
that couple to the Dirac fermions as valley-dependent Dirac masses. 
Similarly, the first proposal for realizing a TMI originated from honeycomb lattice Dirac fermions, which were shown to be dynamically gapped out by finite range density-density interactions~\cite{rag08,wee10,castro11,gru13}. These proposals hold the exciting promise of observing electronically self-organized topological insulators with single-layer graphene as the prime canidate material. Two main complications arise, however, in the case of honeycomb lattice Dirac fermions. 

First, the stability of the QAH state
generally
 relies on physically unrealistic interaction energy scales. In particular, it requires the next-nearest neighbor (NNN) interaction to be stronger than the nearest neighbor (NN) interaction, an unlikely situation in for instance graphene~\cite{weh11}. Second, the vanishing density of states at half filling 
implies, even at zero temperature, 
a finite critical interaction strength for the QAH state 
to be stabilized, 
which also raises 
questions as to the validity of 
the Hartree-Fock approximation.
Recent exact diagonalization studies, indeed,  have not been able to confirm the mean field results~\cite{dag14,gar14}.

In this paper, we present a simple and physically intuitive way to overcome these 
hurdles and realize the TMI on the honeycomb lattice. The central idea of our proposal is to alter the electronic properties of the honeycomb Dirac semimetal by means of substrate-induced electrostatic potentials with an hexagonal superlattice structure of tripled unit cell. 

In their simplest form, these substrate-induced potentials take the form of non-Abelian $SU(2)$ gauge fields in the low-energy descriptions of honeycomb lattice electrons. 
Although gauge potentials of any origin generally shift the Dirac cones in momentum space \cite{vozm10,gop12}, we show that for the 
combinations of ${\cal SU}(2)$ gauge field components originating from an hexagonal underlay with tripled unit cell, the Dirac cones morph into a quadratic band crossing point (QBCP). 
Higher order harmonic components of the superlattice potential respecting translational invariance but making the sublattices inequivalent can remove the QBCP and open up a spectral gap. 
In the QBCP regime, we analyze the effect of interactions in two ways. First, we use a 
perturbative renomalization group (RG) approach to establish that as a consequence of the QBCP the system has a weak-coupling instability 
in much the same way as was established for generic symmetry protected QBCPs~\cite{sun09,dora14}. 
However, in contrast to these models, the possible occurrence of an interaction-induced rotational symmetry-breaking nematic phase ~\cite{sun09,dora14} with the QBCP splitting into two Dirac cones is prohibited by the fact that the hexagonal underlay fully breaks the three-fold rotational symmetry of the honeycomb lattice, leaving the time-reversal symmetry breaking QAH gapped state as the only instability at weak coupling. 
We use Hartree-Fock theory to show that the interaction-induced QAH state is indeed realized at weak coupling.

\begin{figure}[t]
\includegraphics[width=\columnwidth]{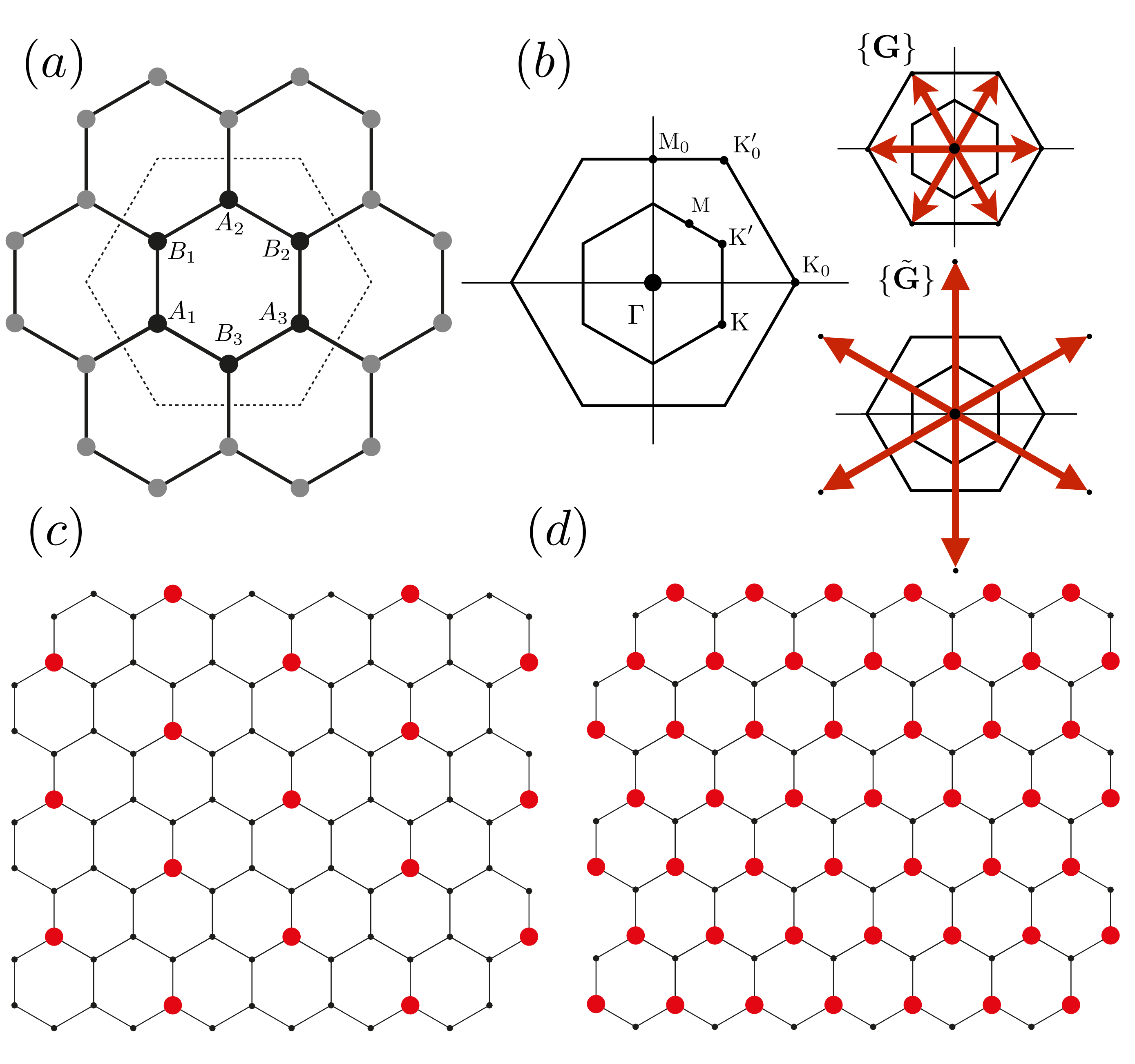}
\caption{(a) Honeycomb lattice with enlarged unit cell. The unit cell, which is marked by the dashed hexagon, contains six sites labeled by $A_\alpha$ and $B_\alpha$ with $\alpha=1,2,3$. (b) On the left the Brillouin zone of graphene and the folded Brillouin zone corresponding unit cell tripling (inner black haxagon). On the right the sets of superlattice wave vectors $\{ {\bf G} \}$ and $\{ {\bf \widetilde{G}} \}$ with respect to the two Brillouin zones. (c) Red dots indicate minima of $\sqrt{3} \times \sqrt{3} $ superlattice potential generated by $\{ {\bf G} \}$ giving rise to the QBCP. (d) Minima of the second order  superlattice components $\{ {\bf \widetilde{G}} \}$. } 
\label{fig:fig1}
\end{figure}

\section{Honeycomb superlattices} 
Superlattices~\cite{tsu} have attracted tremendous interest as they allow to accurately manipulate the band structure of two-dimensional materials and hence hold the promise of tailored electronic properties. 
Superlattices of a Dirac semimetal, with graphene epitaxially grown on prepatterned substrates being the canonical example ~\cite{dea10},
lead to a rich plethora of phenomena such as Dirac fermions cloning \cite{yan12,pon13}  
or strongly anisotropic massless chiral fermions~\cite{par08np}. 
Generally, the physics of these honeycomb superstructures can be described by considering the effect of a substrate-induced external electrostatic potential acting on the pristine honeycomb lattice sites. Within this approach, the generation of secondary Dirac cones with halved group velocity in large graphene Moir\'e superstructures \cite{yan12} has been identified~\cite{ort11,wall12}. We therefore use the same conceptual starting point and consider electrons on a honeycomb lattice, in the presence of an external electrostatic perturbation originating from a commensurate $\sqrt{3} \times \sqrt{3}$ hexagonal underlay, as depicted in Fig.~\ref{fig:fig1}(a). The $\sqrt{3} \times \sqrt{3}$ hexagonal superlattice, having a unit cell three times larger than the elementary honeycomb unit cell, allows for potential configurations leading to a QBCP.

In the most general setting, non-interacting spinless fermions subject to electrostatic potentials with $\sqrt{3} \times \sqrt{3}$ hexagonal periodicity are described by the tight-binding Hamiltonian  
\begin{eqnarray}
{\cal H}_{0}
&=& 
t \sum_{\langle \alpha i,\beta j \rangle} {\hat a}_{\alpha i}^{\dagger} {\hat b}_{\beta j} + {\it h.c.}  \nonumber \\ 
& & 
+\sum_{\alpha=1}^{3} \sum_{i} \left( {\cal V}_{A \alpha} {\hat n}^A_{\alpha i}+  {\cal V}_{B \alpha} {\hat n}^B_{\alpha i}\right)  , 
 \label{eq:tbhamiltonian}
\end{eqnarray}
where $t$ indicates the NN hopping amplitude, the ${\cal V}_{A \alpha}$'s, ${\cal V}_{B \alpha}$'s are the on-site energies renormalized by the substrate perturbation and the fermionic operator $ {\hat a}_{\alpha i}
$ (${\hat b}_{\alpha i} $) annihilates an electron at position $i$ in the
sublattice  $A_{\alpha}$ ($B_{\alpha}$) with $\alpha=1,2,3$. Here $\langle \alpha i,\beta j \rangle$ denotes a sum over all NN combinations of $\alpha i $ and $\beta j$, and ${\hat n}^A_{\alpha i} = {\hat a}_{\alpha i}^{\dagger}{\hat a}_{\alpha i}$ (same for $B$).
As the real space unit cell is tripled, the Brillouin zone (BZ) is folded, with the corners of the hexagonal lattice BZ, the so-called Dirac points, now occurring at the $\Gamma$ point. 

To gain insight in the effect of the substrate-induced potentials on the electronic structure, we consider the effective low-energy theory close to the $\Gamma$ point of the folded Brillouin zone (BZ)  [see the Supplemental Material for details]. The low-energy honeycomb lattice Dirac fermions are modified in the following way 
\begin{eqnarray}  \label{eq:lowenergy}
{\cal H}_{eff}&=& \left[ \Gamma_x \left(v_F  k_x - {\cal A}_x^i {\cal Q}_i \right) + \Gamma _y \left(v_F k_y - {\cal A}_y^i {\cal Q}_i \right)  \right] \nonumber \\ & & +   {\cal V}_+ \,   \tau_0 \sigma_0 +  {\cal V}_- \, \tau_0 \sigma_3, 
\end{eqnarray} 
where  $v_F$ is the Fermi velocity, $\Gamma_x = \tau_3 \sigma_1$, $\Gamma_y = \tau_0 \sigma_2$ and the ${\boldsymbol \sigma}$ and ${\boldsymbol \tau}$ operators  are Pauli matrices acting on the sublattice and valley degrees of freedom of the honeycomb lattice, respectively. In addition, we introduced the Dirac matrices ${\cal Q}_i$ ($i=1,2,3$)~\cite{gop12}, which are given by ${\cal Q}_1= - \tau_2 \sigma_2 $, ${\cal Q}_2 = \tau_1 \sigma_2$, ${\cal Q}_3 = \tau_3 \sigma_0$. These matrices commute
with
the $\Gamma_{x,y}$ matrices, and in addition realize an ${\cal SU}(2)$ pseudo-spin algebra $\left[ {\cal Q}_i , {\cal Q}_j \right] = 2 i \epsilon_{ijk} {\cal Q}_k$. 
We have defined $ {\cal V}_\pm = ({\cal V}_A\pm {\cal V}_B)/2$ as the sum and the difference of the average potentials on each sublattice, i.e. ${\cal V}_{X} \equiv \sum_{i=1}^3 {\cal V}_{Xi}/3$ ($X=A,B$). The sum ${\cal V}_+ $ couples to the identity $\tau_0 \sigma_0$ whereas the difference couples to $\tau_0 \sigma_3$, which anticommutes with the $\Gamma_{x,y}$ and corresponds to an inversion symmetry breaking Dirac mass \cite{ryu09} [c.f.  Fig.~\ref{fig:fig1}(b)].
The remaining four linear combinations of potentials enter as gauge fields ${\cal A}_{x}^{i}$ and ${\cal A}_{y}^{i}$~\cite{gop12} and couple to the ${\cal Q}_i$. The explicit expressions for these linear combinations are summarized in Table~\ref{tab:potentials}. For specific combinations of these pseudo-gauge fields ${\cal A}_{x}^{i}$ and ${\cal A}_{y}^{i}$, the low-energy spectrum becomes quadratic as opposed to Dirac-linear, and these pseudo-gauge field configurations were shown 
to generate an effective nonzero 
non-Abelian
field strength~\cite{juan13}.

Having discussed the general structure of the Hamiltonian, we proceed to show that substrate induced electrostatic potentials can realize such pseudo-gauge field configurations. 
The renormalization of the on-site energies due to a commensurate hexagonal underlay with tripled unit cell can be obtained following the observation \cite{ort11,wall12,wall13} that the electrostatic potential felt by the spinless electrons is smoothened by the large separation between the system and the substrate, as compared to the separation of NN honeycomb lattice sites. We therefore consider a smooth superlattice perturbation with triangular periodicity, expressed as ${\cal V}({\bf r}) = \sum_{\bf G} V_{\bf G} \mathrm{e}^{i {\bf G} \cdot {\bf r}}$. The amplitudes $V_{\bf G}$ only depend on the modulus of ${\bf G}$, and we restrict the ${\bf G}$'s to the simplest set of wavevectors ${\bf G} / G = \left\{\pm 1 , 0 \right \}, \left\{ \pm \cos{\pi /3} , \pm \sin{\pi / 3} \right\}$ \cite{wall13} with equal magnitude $G=4 \pi /  (3 a) $, $a$ being the honeycomb lattice constant. 
We distinguish two alternatives for choosing the origin of the superlattice perturbation ${\cal V}({\bf r})$ with respect to the center of a reference honeycomb lattice hexagon. In case the origin of the superlattice perturbation coincides with the center of the reference hexagon, no symmetries other than translational symmetry are broken and one finds $V_{A i} \equiv V_{B i} \equiv 0$, meaning no electrostatic effect on the electrons at lattice sites. If, however, the center of the superlattice electrostatic potential is aligned with a honeycomb lattice site, the on-site energies on the $A$ sublattice take the values $V_{A1} = 6 V_G$ and $V_{A 2} = V_{A 3} = -3 V_G$, while the $B$-sublattice sites remain unaffected. The latter case is shown in Fig.~\ref{fig:fig1}(c). 
In terms of the effective low-energy Hamiltonian of Eq.~\eqref{eq:lowenergy} this yields the explicit expressions for the gauge fields $A_{x}^{1} \equiv  - A_{y}^{2} \equiv 3 V_G / 2$ and $A_{x}^2 \equiv A_{y}^1 \equiv 0$, while we find that ${\cal V}_+$ and ${\cal V}_-$ vanish (see also Table~\ref{tab:potentials}). Given these expressions, the low-energy dispersion is readily obtained as $E({\bf p}) = \pm  {\cal A}_x^{1}  + \beta  \sqrt{( {\cal A}_x^{ 1})^2 +v_F^2  k^2} $ where $\beta=\pm 1$. Two bands touch at $\Gamma$ to form a QBCP as a direct consequence of the  gauge fields following from the specific arrangement of the substrate.

\begin{table}[t]
\centering
\begin{ruledtabular}
\begin{tabular}{clccc}
 \multicolumn{2}{c}{Matrix} & Potentials $V_{X\alpha}$  & $V_G$ & $ V_{\tilde{G}} $ \\  [1ex]
\hline 
$\tau_0\sigma_0$  && $\mathcal{V_+} = (V_{A}+V_{B})/2$ &- & $3\mathcal{V}_{\tilde{G}}/2$  \\ 
$\tau_0\sigma_3$  & & $\mathcal{V_-} = (V_{A}-V_{B})/2$ & -& $9\mathcal{V}_{\tilde{G}}/2$\\ 
\multirow{2}{*}{$\mathcal{Q}_1=-\tau_2\sigma_2$} &   $\mathcal{A}^1_x $   & $(\text{Re}\,\mathcal{V}^\omega_A-\text{Re}\,\mathcal{V}^\omega_B)/2$  & $3\mathcal{V}_{G}/2$ & - \\ 
 &$\mathcal{A}^1_y $    &   $-(\text{Im}\,\mathcal{V}^\omega_A+\text{Im}\,\mathcal{V}^\omega_B)/2$       &- & - \\ 
\multirow{2}{*}{$\mathcal{Q}_2=\tau_1\sigma_2$} & $ \mathcal{A}^2_x $  &  $-(\text{Im}\,\mathcal{V}^\omega_A-\text{Im}\,\mathcal{V}^\omega_B)/2$  & -& - \\
& $ \mathcal{A}^2_y $  & $(\text{Re}\,\mathcal{V}^\omega_A+\text{Re}\,\mathcal{V}^\omega_B)/2$  &-$3\mathcal{V}_{G}/2$ & -
\end{tabular}
\end{ruledtabular}
 \caption{The effect of substrate induced electrostratic potentials in the low-energy electronic structure. This first column lists the Dirac matrices, the second column the combination of potentials $V_{X\alpha}$ ($X=A,B$) which couple to the respective terms. Note the definitions ${\cal V}_{X} \equiv \sum_{i=1}^3 {\cal V}_{Xi}/3$ and ${\cal V}^\omega_{X} = \left( {\cal V}_{X1} + \omega  {\cal V}_{X2} + \omega^{2}  {\cal V}_{X3} \right) / 3$, where in latter we used $\omega=\exp{[2 \pi i / 3]}$. The third and fourth column list the specific values of these potentials in terms of the potential amplitudes $V_G$ and $V_{\tilde{G}} $, capturing the effect of first and second harmonics, respectively.}
\label{tab:potentials}
\end{table}

Contrary to QBCPs protected by lattice symmetries and carrying a $2 \pi$ Berry flux \cite{sun09,sun11}, the QBCP emerging from the substrate-induced potentials is not protected by any symmetries that may quantize the Berry flux. Topologically stable QBCP can only be gapped out by ${\cal T}$-breaking perturbations, whereas the QBCP engineered by breaking symmetries can be energetically split by ${\cal T}$-invariant perturbations.  Specifically, we find that a full substrate-induced bandgap naturally arises by taking into account the next set of harmonics in the superlattice perturbation ${\cal V}({\bf r})$.  The corresponding wavevectors have equal magnitude ${\widetilde G} = 4 \pi / (\sqrt{3} a)$ and are given by  ${\bf G} / {\widetilde G} =   \left\{0 , \pm 1 \right \}, \left\{ \pm \cos{\pi /6} , \pm \sin{\pi / 6} \right\}$. The inclusion of this additional set of harmonics in ${\cal V}({\bf r})$ does not affect gauge field terms, but rather introduces  a finite $ {\cal V}_-$ which takes the value $  {\cal V}_- = 9 V_{{\widetilde G}} / 2$ and a finite $ {\cal V}_+= 3 V_{{\widetilde G}} / 2$. Apart from the identity term, the dispersion of the low-energy Hamiltonian Eq.~\eqref{eq:lowenergy} then explicitly reads 
$E({\bf p}) = \beta {\cal A}_x^{1}  \pm [v_F^2 k^2 + (A_x^1 + \beta {\cal V}_-)^2]^{1/2}$. With this, massive Dirac fermions and hence a full substrate-induced bandgap occur for $|{\cal A}_x^1| < | {\cal V}_- |$. We find that the transition from the gapless QBCP regime to the gapped one is marked by the presence of a pseudo-spin-1 conical-like spectrum [see the Supplemental Material].

\begin{figure}[t]
\includegraphics[width=\columnwidth]{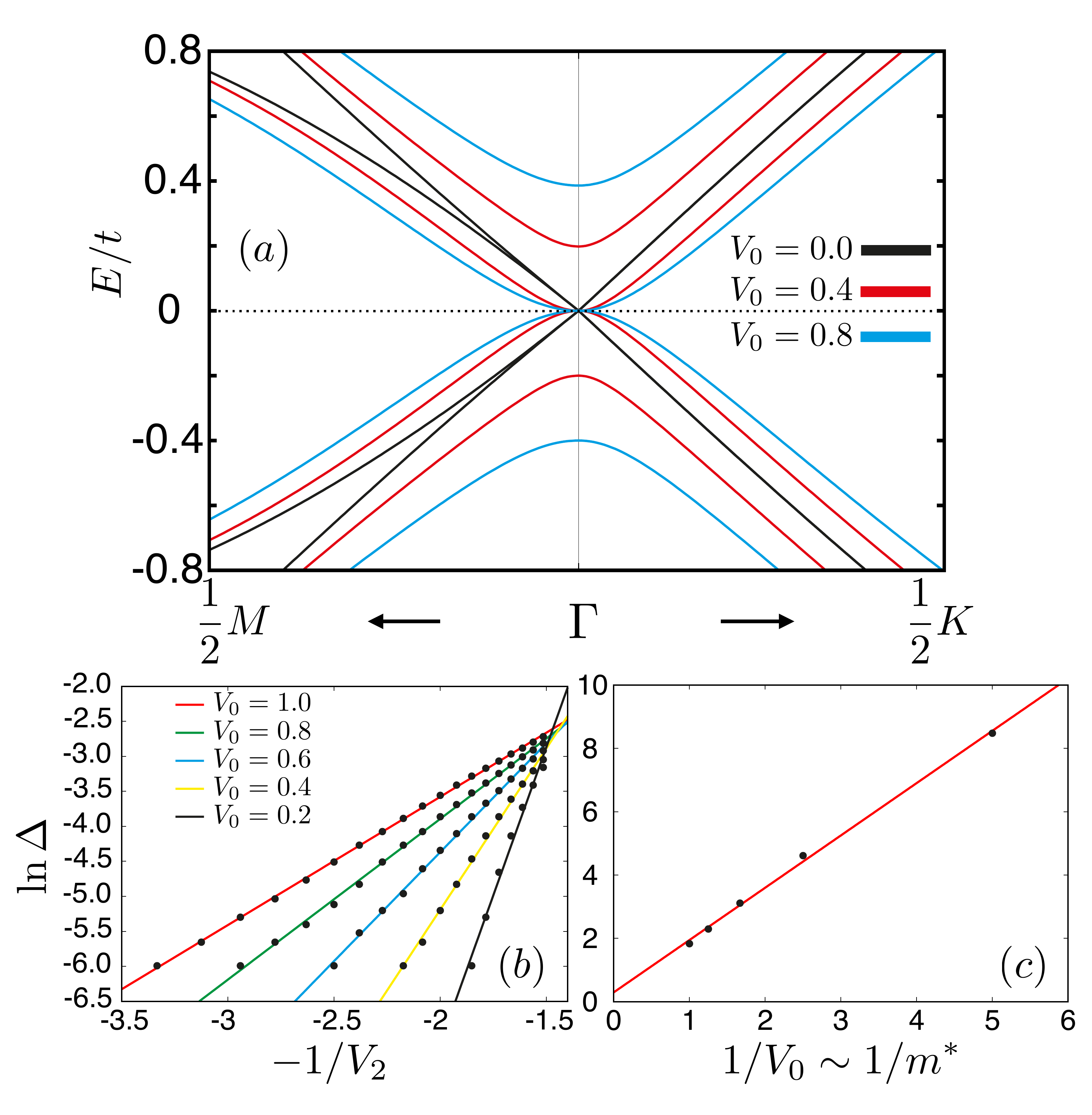}
\caption{(a)  Electronic band structure of graphene in the presence (and absence, $V_0=0.0$) of substrate induced potentials. Representative examples are shown for which the QBCP emerges, i.e. $V_0=0.4$ (red) and $V_0=0.8$ (blue). 
(b) Scaling of the QAH order parameter $\Delta$ as function of the NNN interaction $V_2$ for various values of substrate potentials ($V_0\sim m^*$). Solid lines represent linear fits of $\ln \Delta $ versus $-1/V_2$.
(c) Scaling of the slopes of the linear fits in (b) as funtion of $1/V_0\sim 1/m^*$. Solid line represents a linear fit.  
} 
\label{fig:fig2}
\end{figure}

\section{Interaction-induced \\ topological insulator} 
The possibility to engineer a QBCP in the honeycomb lattice band structure at half filling using a hexagonal superlattice 
suggests a closer investigation of electron-electron interactions and their effect on the electronic ordering in the QBCP regime. The effect of interactions on a QBCP have been studied previously both with RG methods and mean field theory~\cite{sun09,dora14}, but since 
the QBCP under consideration here is different in nature, these results do not directly apply.
In order to establish whether the electrostatic potential-induced QBCP is still marginally unstable to weak repulsive interactions, we have employed a perturbative RG approach.
To this end, we first obtained a continuum model of spinless interacting electrons on the honeycomb lattice. The non-interacting continuum theory contains the substrate potentials and we only retain the first set of harmonics setting $V_0 = V_{A1} = 6V_G$.  To obtain a continuum QBCP theory we take the Dirac Hamiltonian of Eq.~\eqref{eq:lowenergy} and project the momentum dependent part into the two-component low-energy subspace at $\Gamma$. We find that the substrate potentials enter as an effective mass, i.e. $m^* \equiv V_0\hbar^2/(4v_F^2)$, and hence control the density of states (DOS). There are two interactions to consider, the NN interaction $V_1$ and NNN interaction $V_2$ given by the Hamiltonian $H_{V_1V_2} =  V_{1} \sum_{\langle i,j \rangle} \hat{n}_i \hat{n}_j + V_{2} \sum_{ \langle \langle i,j \rangle \rangle} \hat{n}_i \hat{n}_j $. Deriving the effective continuum vertices, i.e. projecting the interactions into the low-energy subspace, shows that $V_1$ is irrelevant in the weak coupling regime~\cite{note1}
due to the specific structure of the low-energy states. These are localized exclusively on one of the sublattices, the $B$ sublattice for our choice of potentials, and hence an intersublattice interaction cannot contribute. In deriving the RG $\beta$-function we follow the scheme laid out in~\cite{shan94}. We find,
that to one-loop order, the RG-$\beta$ function is given by
$\beta(V_2) = \partial V_2 /\partial \log s = \alpha V^2_2$ with $\alpha = | V_0 | / (16 \pi v_F^2) =| m^* | / (4 \pi \hbar^2)$, which is equivalent in structure to the result obtained in~\cite{sun09,dora14} and we thus conclude that the coupling $V_2$ flows to strong coupling.

Based on the result that $V_2$ flows to strong coupling, we employ mean-field techniques to determine the type of ordering that is realized. As a first step, we have calculated the normal state susceptibilities $\chi$ to various orders in the absence of potentials, and find that fluctuations in the QAH channel are strongest (see the Supplemental Material). In the presence of substrate potentials, which engineer the weak-coupling instability, one thus expects the interaction-induced QAH state. This is confirmed by extensive restricted and unrestricted mean-field calculations, performed for a range of parameters $(V_0,V_1,V_2)$ at zero temperature. In the mean-field calculations we have explicitly allowed for the formation of intra-sublattice charge redistribution, as these have lower energy than the QAH state at large $V_2$ in case of pristine graphene~\cite{gru13}. Details of the mean-field decoupling in the six-atom unit cell may be found in Ref.~\onlinecite{gru13}. For finite $V_2$ we consistently find the QAH state as the mean-field ground state.

The RG calculation provides us with quantitative predictions regarding the scaling of the QAH order parameter $\Delta$ as function of coupling constant $V_2$. We have used restricted mean field calculations, i.e. only decoupling in the QAH channel, to check these predictions. Specifically, one expects that $\Delta $ scales as $ \Delta \sim \Lambda e^{-c/(m^*V_2)}$ where $\Lambda$ is an energy cutoff of the order of the bandwidth, and $c= 8 \pi \hbar^2$. Hence, we expect $\ln \Delta$ to depend linearly on $-1/V_2$. Fig.~\ref{fig:fig2}(b) shows a linear fit of $\ln \Delta$ obtained from numerical restricted mean field calculations. We observe that the linear fit works well for values of the superlattice potentials ranging from $V_0=0.2$ to $V_0=1.0$. Fig. ~\ref{fig:fig2}(c) shows a linear fit of the slope of the linear fits of Fig.~\ref{fig:fig2}(b) as function of $1/V_0 \sim 1/m^*$. Again one expects linear behaviour which the panel (c) clearly shows. Based on both the RG and the mean field approaches we therefore conclude that the substrate induced QBCP gives rise to a weak-coupling instability towards a time-reversal breaking QAH state. 

\section{Conclusions}
We have shown, in conclusion, the emergence of a QBCP in the half-filled honeycomb lattice resulting from electrostatic coupling to a substrate with hexagonal symmetry but with $\sqrt{3} \times \sqrt{3}$ periodicity, i.e. a tripled unit cell. The superlattice potential couples to the low-energy Dirac fermions as a specific linear combinations of pseudo gauge fields of $SU(2)$ type, corresponding to a nonzero non-Abelian field strength. 
The QBCP we have shown to arise in the presence of hexagonal superlattices carries a trivial zero Berry flux and can be removed in favor of a full spectral gap by  additional 
modulations of electrostatic potentials. 
In the QBCP regime, we have shown that a topological quantum anomalous Hall phase can be generated by repulsive NN and NNN interactions even when the latter is small, a regime which is naturally realized in graphene. 
A one-loop RG analysis supplemented by Hartree-Fock mean field calculations demonstrate that the quadratic low-energy dispersion 
is marginally unstable to the formation of the QAH phase at arbitrarily weak repulsive electronic interactions. 

Using density functional theory (DFT) a QBCP has been recently found in a graphene-indium chalcogenide heterostructure where single layer graphene is deposited on top of hexagonal In$_2$Te$_2$ monolayers \cite{gio15}. For this prototypical bilayer, the DFT characteristic strength of the electrostatic modulated potential $V_0 \simeq 0.2 \, t$, for which   
our Hartree-Fock mean-field calculations predict  the QAH gap to reach room temperature at an effective N.N.N. interaction $V_{2}  \simeq 0.58 t$, a value smaller than the effective Coulomb interaction of single-layer graphene \cite{weh11}. This observation suggests that the interaction-driven QAH state can be realized in the experimental realm using materials such as PtTe$_2$, h-GaTe \cite{wall13}, and h-InSe as graphene substrates. 

\section{Acknowledgements}
We gratefully acknowledge interesting discussions with M. Daghofer and G. Giovannetti. J.V. acknowledges support from the Netherlands Organization of Scientific Research. C.O. acknowledges the financial support of the Future and Emerging Technologies (FET) programme within the Seventh Framework Programme for Research of the European Commission, under FET-Open grant number: 618083 (CNTQC). This work has been supported by the Deutsche Forschungsgemeinschaft under Grant No. OR 404/1-1 and SFB 1143.


\begin{thebibliography}{10}
\bibitem{note1} At tree level, nearest-neighbor interactions are irrelevant in the RG sense since the corresponding vertex is proportional to $k^2$.
\end{thebibliography}
\end{document}